\documentclass[twocolumn,trackchanges]{aastex631} 

\newcommand{\nii}{[\ion{N}{2}] $\lambda$6584}
\newcommand{\oiii}{[\ion{O}{3}] $\lambda$5007}
\newcommand{\sii}{[\ion{S}{2}] $\lambda\lambda$6717, 6731}

\usepackage{booktabs}

\begin{document}

\title{Discovery of a Pair of Galaxies with Both Hosting X-ray Binary Candidates at $z=2.544$}

\correspondingauthor{Zheng Cai}
\email{zcai@mail.tsinghua.edu.cn}

\author[0009-0003-4133-0292]{Sijia Cai}
\affiliation{Department of Astronomy, Tsinghua University, Beijing 100084, China}

\author[0000-0001-8467-6478]{Zheng Cai}
\affiliation{Department of Astronomy, Tsinghua University, Beijing 100084, China}

\author[0000-0002-6221-1829]{Jianwei Lyu}
\affiliation{Steward Observatory, University of Arizona, 933 North Cherry Avenue, Tucson, AZ 85721, USA}

\author[0000-0003-0111-8249]{Yunjing Wu}
\affiliation{Department of Astronomy, Tsinghua University, Beijing 100084, China}

\author[0000-0001-6052-4234]{Xiaojing Lin}
\affiliation{Department of Astronomy, Tsinghua University, Beijing 100084, China}

\author[0000-0001-6251-649X]{Mingyu Li}
\affiliation{Department of Astronomy, Tsinghua University, Beijing 100084, China}

\author[0000-0001-7557-9713]{Junjie Mao}
\affiliation{Department of Astronomy, Tsinghua University, Beijing 100084, China}

\author[0000-0001-7829-7797]{Jiayi Chen}
\affiliation{Department of Astronomy, Tsinghua University, Beijing 100084, China}

\author[0009-0000-4701-4934]{Pengjun Lu}
\affiliation{Department of Astronomy, Tsinghua University, Beijing 100084, China}



\begin{abstract}

Among high-redshift galaxies, aside from active galactic nuclei (AGNs), X-ray binaries (XRBs) can be significant sources of X-ray emission. XRBs play a crucial role in galaxy evolution, reflecting the stellar populations of galaxies and regulating star formation through feedback, thereby shaping galaxy structure. In this study, we report a spectroscopically confirmed X-ray emitting galaxy pair (UDF3 and UDF3-2) at $z = 2.544$. By combining multi-wavelength observations from JWST/NIRSpec MSA spectra, JWST/NIRCam and MIRI imaging, Chandra, HST, VLT, ALMA, and VLA, we analyze the ionized emission lines, which are primarily driven by H II region-like processes. Additionally, we find that the mid-infrared radiation can be fully attributed to dust emission from galaxy themselves. Our results indicate that the X-ray emission from these two galaxies is dominated by high-mass XRBs, with luminosities of 
$L_X=$ \((1.43\pm0.40) \times 10^{42} \, \text{erg} \, \text{s}^{-1}\) for UDF3, and \((0.40\pm0.12) \times 10^{42} \, \text{erg} \, \text{s}^{-1}\) for UDF3-2. Furthermore, we measure the star formation rate (SFR) of $529_{-88}^{+64}$ $M_\odot$ yr$^{-1}$ for UDF3, placing it $\approx$ 0.5 dex below the $L_X$/SFR-$z$ relation. This offset reflects the redshift-dependent enhancement of $L_X$/SFR-$z$ relation, which is influenced by metallicity and serves as a key observable for XRB evolution. In contrast, UDF3-2, with the SFR of $34_{-6}^{+6}$ $M_\odot$ yr$^{-1}$, aligns well with the $L_X$/SFR-$z$ relation. This galaxy pair represents the highest-redshift non-AGN-dominated galaxies with individual X-ray detections reported to date. This finding suggests that the contribution of XRBs to galaxy X-ray emission at high redshift may be underestimated.

\end{abstract}

 

\section{Introduction} \label{sec:intro}

X-ray binaries (XRBs) play a crucial role in galaxy evolution. They provide insights into the underlying stellar population and also act as active agents in shaping their host galaxies through energetic feedback (e.g. \citealt{belczynski2004x}; \citealt{gallo2005dark}; \citealt{tetarenko2018mapping}; \citealt{tetarenko2020jet}; \citealt{misra2023x}).

X-ray emission from XRBs is a powerful tracer of star forming activities in galaxies (\citealt{grimm2003high}; \citealt{ranalli20032}; \citealt{cowie2012faintest}). In galaxies without significant AGN activity, XRBs, composed of a compact object and a stellar companion, dominate X-ray emission above 1--2 keV through accretion from the companion \citep{gilfanov2023x}. Studies of nearby galaxies show that XRB populations scale with the star formation rate (SFR) and stellar mass ($M_\star$) of their host galaxies \citep{gilfanov2004low, mineo2012x, zhang2012dependence}. High-mass X-ray binaries (HMXBs), linked to massive stars, emit X-rays within 10–100 Myr of a starburst, with X-ray luminosity ($L_X$) closely tied to SFR \citep{iben1995model, bodaghee2011clustering, antoniou2016star}. In contrast, low-mass X-ray binaries (LMXBs), associated with low-mass stars in binary systems that evolve on timescales of about 1-10 Gyr, trace the integrated star formation history of the galaxy - and consequently its total stellar mass \citep{lewin1997x}. In galaxies with specific SFR (sSFR) greater than $10^{-10} \, \text{yr}^{-1}$, HMXBs could dominate the X-ray emission (\citealt{lehmer2010chandra}; \citealt{lehmer2019x}).

Population synthesis models predict that metallicity is also a key factor shaping the HMXB population, with lower metallicity environments leading to higher X-ray luminosities (\citealt{linden2010effect}; \citealt{fragos2013x}; \citealt{tremmel2013modeling}). In such environments, weaker stellar winds driven by radiation pressure on atomic lines—due to reduced UV absorption by metals—result in more massive stars in binaries, with less angular momentum loss and tighter orbits, thereby increasing the likelihood of Roche lobe overflow and the formation of more luminous HMXBs. Models predict a positive evolution of the $L_X$/SFR ratio with redshift. With the advent of deep extragalactic X-ray surveys, observations have roughly found that $L_X$/SFR $\propto$ (1+$z$) (\citealt{basu2012x}; \citealt{lehmer2016evolution}; \citealt{aird2016x}). Until recently, stacking of X-ray images from optically selected galaxies at $z \sim 2.5$ has provided evidence for a correlation between $L_X$/SFR evolution and metallicity (\citealt{fornasini2019mosdef}), which is consistent with similar findings in nearby galaxies (\citealt{fornasini2020connecting}; \citealt{lehmer2024empirical}), thereby supporting the above physical picture. These studies may offer stronger constraints to explore the potential role of XRBs in heating the intergalactic medium during the epoch of reionization, when the universe was extremely metal-poor (\citealt{dijkstra2012constraints}; \citealt{artale2015stellar}; \citealt{das2017high}; \citealt{madau2017radiation}).

Therefore, studying X-ray emission from star-forming galaxies with no AGN dominating at high redshift is crucial for understanding XRB evolution in the context of cosmic star formation and refining $L_X$ as an SFR indicator. 
The James Webb Space Telescope (JWST), with its wide-band imaging and spectroscopy, is revolutionizing our understanding of cosmic star formation and metal enrichment history (e.g. \citealt{ferrara2023stunning}; \citealt{shapley2023jwst}; \citealt{wang2024rubies}; \citealt{lyu2024active}; \citealt{llerena2024ionizing}). JWST is poised to aid in studying the collective properties of XRB populations at high redshifts, especially when combined with Deep X-ray observations, providing an unprecedented opportunity to precisely select high-$z$ X-ray detected galaxy samples while distinguishing them from those with AGN contributions. This will offer new insights into their role in galaxy evolution during earlier cosmic epochs.

In this study, we search for Chandra X-ray detected star-forming galaxies in the GOODS-S field. By combining observations from the VLA and ALMA, spectroscopic data from JWST/NIRSpec and VLT/VIMOS, and photometry from HST, JWST/NIRCam and MIRI, we identify a pair of galaxies at $z_\mathrm{spec} = 2.544$ (UDF3 and UDF3-2). From the emission line diagnostics, full-spectra and multiple-wavelength SED fitting from rest-frame UV to mid-infrared, as well as morphological analysis, we confirm that the X-ray emission from both member galaxies in this pair are contributed by X-ray binaries and should not be dominated by AGN, making this system hosting the highest redshift X-ray binaries with individual X-ray detection identified so far. 

The paper is structured as follows: Section \ref{sec:data} introduces the data and our targets; Section \ref{sec:analysis} presents the analysis of the spectra and photometry; Section \ref{sec:results} reports the results and discussion; and Section \ref{sec:conclusion} concludes the study. We adopt a cosmological model with $H_0 = 67.8 \, \mathrm{km\, s^{-1} \, Mpc^{-1}}$, $\Omega_M = 0.31$, and $\Omega_\Lambda = 0.69$ (Planck Collaboration et al. \citeyear{ade2016planck}).

\section{Data and Target} \label{sec:data}

\begin{figure*}[htbp]
\centering
\includegraphics[width=\textwidth]{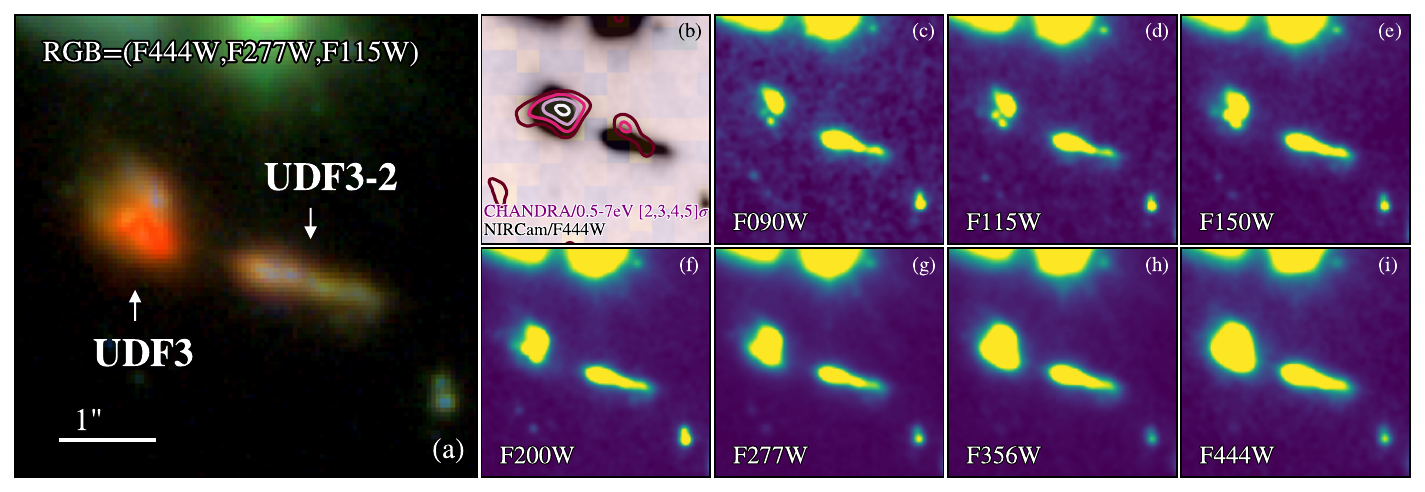} 
\caption{RGB and multi-band images of galaxies UDF3 and UDF3-2, each with a cutout size of $5'' \times 5''$ and a pixel scale of $0.05''$/pixel. (a) RGB composite image created using the F444W, F277W, and F115W filters. (b) Full-band (0.5--7 keV) Chandra X-ray image ($0.5''$/pixel) with [2,3,4,5]$\sigma$ brightness contours overlaid on the JWST/NIRCam F444W filter image. (c--i) NIRCam images in various filters, each smoothed with a 1$\sigma$ Gaussian kernel.}
\label{fig:images}
\end{figure*}

\begin{figure*}[htbp]
\centering
\includegraphics[width=\textwidth]{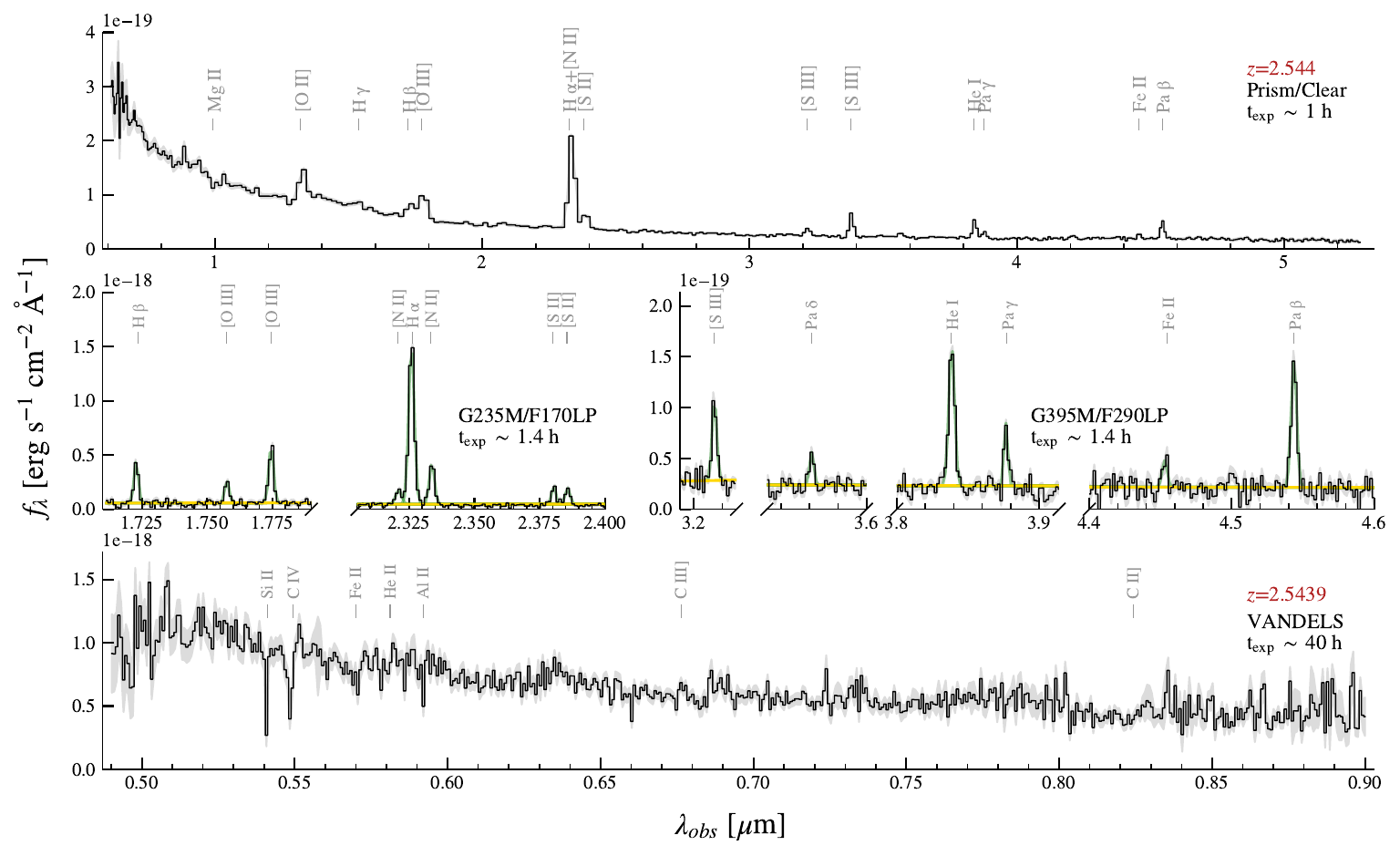} 
\caption{
The top three panels show the NIRSpec spectra of UDF3 from the DJA dataset, with the upper panel showing the prism spectrum, and the middle two panels displaying the grating spectra from G235M/F170LP and G395M/F290LP. We only show the emission line regions for clearer visualization of the emission line flux measurements described in Section \ref{sec:Emission-line Fluxes}. The yellow curve represents the power-law fit to the continuum, and the green curve shows the Gaussian fit to the emission lines. The bottom panel presents the VLT/VIMOS spectrum of UDF3-2 from the VANDELS survey. We bin the spectrum over 3 pixels to enhance the \textit{S/N}. The solid lines show the flux density, the shaded regions show the corresponding 1$\sigma$ uncertainties, and the emission and absorption lines are labeled.}
\label{fig:specs}
\end{figure*}

\subsection{Data}

We collected multi-band and survey data in the GOODS-S field. For photometric analysis, we rely on ultraviolet to near-infrared data from catalog released by the JWST Advanced Deep Extragalactic Survey (JADES\footnote{\url{https://archive.stsci.edu/hlsp/jades}}, \citealp{eisenstein2023overview}), which includes HST ACS/WFC filters (F435W, F606W, F775W, F814W, F850LP), HST WFC3/IR filters (F105W, F125W, F140W, F160W), as well as broad- and medium-band filters from JWST/NIRCam (F090W, F115W, F150W, F200W, F277W, F356W, F444W; F182M, F210M, F335M, F410M, F430M, F460M, F480M). In addition, we incorporate the photometric catalog from the Systematic Mid-infrared Instrument (MIRI) Legacy Extragalactic Survey (SMILES\footnote{\url{https://archive.stsci.edu/hlsp/smiles}}, \citealp{rieke2024smiles}), which includes JWST/MIRI filters (F560W, F770W, F1000W, F1280W, F1500W, F1800W, F2100W, F2550W). These two surveys used PSF-matched Kron apertures. The detailed data processing and photometry methods are described in the data release papers of JADES \citep{rieke2023jades} and SMILES \citep{alberts2024smiles}, while the photometry results are listed in the Appendix.

For spectroscopy, we use NIRSpec/MSA spectra from the DAWN \textit{JWST} Archive (DJA\footnote{\url{https://dawn-cph.github.io/dja/spectroscopy/nirspec/}, \url{https://github.com/gbrammer/msaexp}}, \citealp{heintz2024jwst}), which includes low-resolution prism spectra (R $\sim$ 100) and three medium-resolution grating spectra (G140M, G235M, G395M; R $\sim$ 1000) from JADES, along with reduced 1D and 2D spectra. Furthermore, ultra-deep spectra obtained with the VLT/VIMOS medium-resolution (MR) grism in the Chandra Deep Field South (CDFS) from the VANDELS survey \citep{garilli2021vandels} are employed to refine the accuracy of the spectroscopic redshifts.

To investigate X-ray emission, we crossmatch the X-ray source catalog from \citet{luo2016chandra}, along with background-subtracted images, which combine data from approximately 7 Ms of Chandra exposure in the CDFS field. These observations span three energy bands: 0.5–7.0 keV, 0.5–2.0 keV, and 2–7 keV. This catalog represents the deepest and highest-resolution X-ray survey to date, providing invaluable insights into high-$z$ sources.

\subsection{Identification of our target, an galaxy pair at \textit{z}=2.544}

Through cross-matching, we identify a galaxy pair at \textit{z}=2.544. The multi-band images are presented in Figure \ref{fig:images} and the spectra are shown in Figure \ref{fig:specs}.

The source on the left, UDF3, (53.160625, -27.776269), is initially identified in the Hubble Ultra Deep Field (HUDF). It has since been subject to follow-up observations in multiple wavelengths. In the MUSE HUDF observations \citep{bacon2017muse}, its redshift was confirmed to be 2.541. At radio band, star-forming galaxies in some HUDF fields, including UDF3, have had their 5 cm radio SFR surface density estimated using VLA observations, 
consistent with the distribution of stellar-mass surface density estimated from spatially resolved SED fitting using HST images \citep{rujopakarn2016vla}. Further observations have been conducted by ALMA \citep[e.g.][]{dunlop2017deep,franco2018goods,cowie2018submillimeter,buat2019cold,pantoni2021unveiling,pantoni2021alma,heintz2024jwst}. \citet{pantoni2021alma} detected a double-peaked CO (3-2) line, and in conjunction with velocity and velocity dispersion maps, they classified UDF3 as a dusty star-forming galaxy with a tilted disc of rotating molecular gas. Using NIRSpec spectra, we confirm the redshift as \textit{z}=2.544 through the emission lines described in Section \ref{sec:Emission-line Fluxes}.

The source on the right, located at (53.160077, -27.776356), is referred to as UDF3-2 in the following sections. Based on 40 hours exposure from the VANDELS survey, with a spectral resolution of R $\sim$ 650, we detect \ion{Si}{2} $\lambda$1526 and \ion{C}{4} $\lambda\lambda$1548, 1550 absorption lines with \textit{S/N} $>$ 2, and then determine the redshift of UDF3-2 to be 2.5439, which is consistent with that of UDF3.

With the deep Chandra X-ray observation, both sources show $>3\sigma$ detections in full-band images (see Figure \ref{fig:images}~b). Due to the use of an extraction region that approximates the $\approx$ 90\% enclosed counts fraction (ECF) contour of the local PSF for photometry, we consider this to encompass the total X-ray detection from both sources which have a 0.5--7 keV flux of $4.45 \times 10^{-17}$ erg cm$^{-2}$ s$^{-1}$, and a 0.5--2 keV flux of $3.08 \times 10^{-17}$ erg cm$^{-2}$ s$^{-1}$. The 90\% confidence-level upper limit of 2--7 keV flux is $3.57 \times 10^{-17}$ erg cm$^{-2}$ s$^{-1}$. The intrinsic 0.5--7.0 keV luminosity is $L_{X,\text{int}} = 4.55^{+1.41}_{-1.22} \times 10^{42}$ erg s$^{-1}$, with an effective power-law photon index of $\Gamma = 2.44$. 

To deblend the two sources and estimate their intrinsic X-ray luminosities, we assume a power-law SED, where the 2--10 keV luminosity is derived from the 0.5--7 keV luminosity and photon index above. We then perform forward 2D Gaussian modeling on the images of the two sources, resulting in the following luminosities: \( L_{2-10 \, \text{keV}, \, \text{UDF3}} = (1.43\pm0.40) \times 10^{42} \, \text{erg s$^{-1}$} \) and \( L_{2-10 \, \text{keV}, \, \text{UDF3-2}} = (0.40\pm0.12) \times 10^{42} \, \text{erg s$^{-1}$} \). Unless otherwise specified, all \( L_X \) mentioned below refer to the 2--10 keV luminosity.

\section{Analysis} \label{sec:analysis}

\subsection{Emission-line Fluxes} \label{sec:Emission-line Fluxes}

Collisionally excited emission lines trace gas-phase metallicity evolution, ionization state, dust extinction, and dominant excitation sources in galaxies. Thanks to the NIRSpec spectra, we are able to use grating data with R $\sim$ 1000 to resolve adjacent emission lines like H$\alpha$ and [\ion{N}{2}] $\lambda$6584. After masking the emission line regions, we fit the continuum via a power-law function, followed by Gaussian fitting of the emission lines to measure the fluxes of lines. We obtain emission line measurements with \textit{S/N} $>$ 2. The measured line fluxes are $f_{\text{H}\alpha} = (4.00 \pm 0.09)\times10^{-17}$ erg cm$^{-2}$ s$^{-1}$,  $f_{\text{H}\beta} = (8.10 \pm 0.84)\times10^{-18}$ erg cm$^{-2}$ s$^{-1}$, $f_{\text{[OIII]}\lambda5007} = (12.30\pm0.83)\times10^{-18}$ erg cm$^{-2}$ s$^{-1}$, $f_{\text{[NII]}\lambda6584} = (9.41\pm0.58)\times10^{-18}$ erg cm$^{-2}$ s$^{-1}$, $f_{\text{[SII]}\lambda6716} = (3.26\pm0.46)\times10^{-18}$ erg cm$^{-2}$ s$^{-1}$, and $f_{\text{[SII]}\lambda6731} = (3.35\pm0.45)\times10^{-18}$ erg cm$^{-2}$ s$^{-1}$ (also see in Appendix). Under the assumption of Case B recombination , where the theoretical H$\alpha$/H$\beta$ ratio is 2.86, the Balmer decrement analysis gives 
an optical extinction of $A_{\text{V}} = 1.9 \pm 0.4$ for UDF3 .

\subsection{Gas-phase Metallicity} \label{sec:Gas-phase Metallicity}

To fairly compare with previous studies \citep[e.g.][]{fornasini2019mosdef,brorby2016enhanced}, we adopt the \citet{pettini2004iii} calibration to estimate the gas-phase oxygen abundance (O/H) as a proxy for the metallicity of the young stellar population in galaxies, given by:
\begin{equation}
12 + \log(\text{O/H}) = 8.73 - 0.32 \times \text{O3N2}
\end{equation}
\begin{equation}
\text{O3N2} = \log \left( \frac{\text{[\ion{O}{3}]} ~\lambda5007 / \text{H}\beta}{\text{[\ion{N}{2}]} ~\lambda 6584 / \text{H}\alpha} \right)
\end{equation}
where the O3N2 indicator is derived from the fluxes of emission lines near sites of ionizing sources. For UDF3, the derived metallicity is $12 + \log(\text{O/H}) = 8.13 \pm 0.05$.

\subsection{Spectral Energy Distribution Modeling}

\begin{table*}[htpb]
\centering
\caption{Parameter Setup of the \texttt{Prospector} Code and Results after SED fitting}
\label{tab:sed}
\begin{tabular}{lcllcc}
\toprule
\toprule
Parameter & Prior/Value & Unit & Description & \multicolumn{2}{c}{Result} \\
\cmidrule(lr){5-6}
          &              &     &         & UDF3    & UDF3-2 \\
\midrule
\multicolumn{6}{c}{Stellar Component} \\
\midrule
zred & 2.544 & dimensionless & known spectroscopic redshift & - & -\\
mass & [1e8, 1e12], LogUniform & $M_\odot$ & formed stellar mass & $9.8_{-0.1}^{+0.1}$ & $9.6_{-0.1}^{+0.1}$ \\
logzol & [-2, 0.19], TopHat & log($Z/Z_\odot$) & stellar metallicity & $0.16_{-0.07}^{+0.02}$ & $-0.18_{-0.11}^{+0.13}$ \\
sfh & 4 & dimensionless & delay-tau SFH & - & -\\
tage & [0.001, 2.6], TopHat & Gyr & stellar age & $0.015_{-0.001}^{+0.003}$ & $0.206_{-0.066}^{+0.134}$ \\
tau & [0.1, 30], LogUniform & Gyr$^{-1}$ & e-folding time of the SFH & $0.3_{-0.8}^{+0.8}$ & $0.3_{-0.8}^{+0.8}$ \\
dust\_type & 4 & dimensionless & \citet{kriek2013dust} attenuation law & - & -\\
dust2 & [0, 4], TopHat & dimensionless & stellar optical depth at 5500\AA & $2.0_{-0.1}^{+0.1}$ & $0.2_{-0.1}^{+0.1}$  \\
dust\_index & [-0.6, 0.3], TopHat & dimensionless & power-law index of the
attenuation curve & $0.2_{-0.1}^{+0.1}$ & $-0.3_{-0.1}^{+0.1}$ \\
\midrule
\multicolumn{6}{c}{Dust Emission \citep{draine2007infrared}} \\
\midrule
duste\_umin & [0.1, 25], LogUniform & MW value & minimum starlight strength & $1.0_{-0.3}^{+0.2}$ & $0.2_{-0.8}^{+0.8}$ \\
duste\_qpah & [0, 10], TopHat &  \% & fraction of grain mass in PAHs & $1.3_{-0.3}^{+0.5}$ & $0.6_{-0.4}^{+0.5}$ \\
duste\_gamma & [1e-4, 1], LogUniform & dimensionless & fraction of dust heated by starlight at U$>$U$_{\text{min}}$ & $-0.2_{-0.3}^{+0.2}$ & $-2.2_{-1.2}^{+0.9}$ \\
\midrule
\multicolumn{6}{c}{AGN Component \citep{nenkova2008agn}} \\
\midrule
fagn & [1e-15, 1], LogUniform & dimensionless & L$_{\text{AGN}}$/L$_{\text{bol,stellar}}$ & $-6.2_{-0.5}^{+0.3}$ & $-8.6_{-4.3}^{+4.3}$ \\
agn\_tau & [0.1, 150], LogUniform & dimensionless & optical depth of the AGN dust torus & $0.6_{-1.1}^{+1.1}$ & $0.6_{-1.1}^{+1.1}$ \\
\midrule
\multicolumn{6}{c}{Estimated Parameters} \\
\midrule
log($M_\star$/$M_\odot$) & -  & dimensionless & surviving stellar mass & $9.8_{-0.1}^{+0.1}$  & $9.5_{-0.1}^{+0.1}$\\
SFR & - & $M_\odot$ yr$^{-1}$ & star formation rate & $529_{-88}^{+64}$ & $34_{-6}^{+6}$\\
log(sSFR) & - & yr$^{-1}$ & specific star formation rate & $-7.0_{-0.1}^{+0.1}$ & $-8.0_{-0.2}^{+0.2}$\\
\bottomrule
\end{tabular}
\tablecomments{For all parameters whose priors are in LogUniform format, the results are given on a log10 scale.}
\end{table*}

We perform SED fitting to resolve the components and derive the SFR and $M_\star$ of the sources. To minimize the number of free parameters and reliably estimate the contribution of each component in the source, we use the standard \texttt{Prospector} code \citep{johnson2021stellar} with the \texttt{FSPS} model (\citealt{conroy2009propagation}; \citealt{conroy2010propagation}) for modeling. For the stellar component, we assume a \citet{kroupa2001variation} initial mass function and a delayed-$\tau$ star formation history (SFH) described as:
\begin{equation}
    \text{SFR}(t) = A * t e^{-t/\tau}
\end{equation}
\begin{equation}
    A = \frac{M_{\star,\text{formed}}}{\int t e^{-t/\tau} dt}
\end{equation}
where A is the normalization factor.
The dust attenuation is modeled using the \citet{kriek2013dust} dust attenuation curve, which includes a flexible Calzetti curve slope and a UV bump at 2175 \AA, tailored to accommodate the high redshift of our sources. In addition, we incorporate the \citet{draine2007infrared} dust emission model, which assumes that dust grains with a particle size distribution (described by the fraction of dust mass in PAHs, i.e. Q$_{\text{pah}}$) are exposed to a radiation field with a power-law distribution of starlight intensity, where the minimum starlight intensity is \( U_{\text{min}} \). In this model, dust with the mass fraction of \( \gamma \) located in photodissociation regions near luminous stars is heated by starlight with \( U > U_{\text{min}} \) to high temperatures, while the remaining dust is exposed to starlight with \( U = U_{\text{min}} \), remaining at lower temperatures. For the AGN component, we adopt the templates from \citet{nenkova2008agn} to model the dust torus. Since the redshift is fixed, a total of 11 free parameters are included in the fitting (see Table \ref{tab:sed}): (1) formed stellar mass (\texttt{mass}), (2) stellar metallicity (\texttt{logzsol}), (3) stellar age (\texttt{tage}), (4) the e-folding time of the star formation history (\texttt{tau}), (5) attenuation level at 5500 \AA\ (\texttt{dust2}), (6) the power-law index of the attenuation curve (\texttt{dust\_index}), (7) the minimum starlight intensity (\texttt{duste\_umin}), (8) the fraction of grain mass in PAHs (\texttt{duste\_qpah}), (9) the fraction of dust heated at $U>U_{\text{min}}$ (\texttt{duste\_gamma}), (10) fraction of bolometric luminosity between AGN and stellar (\texttt{fagn}), and (11) the optical depth of the AGN dust torus (\texttt{agn\_tau}). As input for the SED fitting, we use PSF-convolved Kron photometry from the JADES catalog, and for the SMILES catalog, aperture corrections have already been applied to account for PSF variations across filters. To account for residual uncertainties in the photometry and limitations in the modeling, we additionally apply a $5\%$ systematic error floor to all bands during SED fitting.
We correct the photometry of UDF3 by subtracting the contributions from all emission lines present in the spectrum shown in the top panel of Figure \ref{fig:specs}. Bayesian posterior sampling is performed using the dynamic nested sampler \texttt{dynesty} \citep{speagle2020dynesty}. We report the median of the posterior distribution, with 1$\sigma$ error bars corresponding to the 16th and 84th percentiles. We report the SFRs averaged over the most recent 10 Myr. We derive the surviving stellar mass based on the predicted mass fractions extracted from the model and subsequently determine the sSFR.

\section{Results and Disscusion} \label{sec:results}

\subsection{The BPT Diagram Indicates an Absence of AGN Component} 

\begin{figure}
    \centering
    \includegraphics[width=\linewidth]{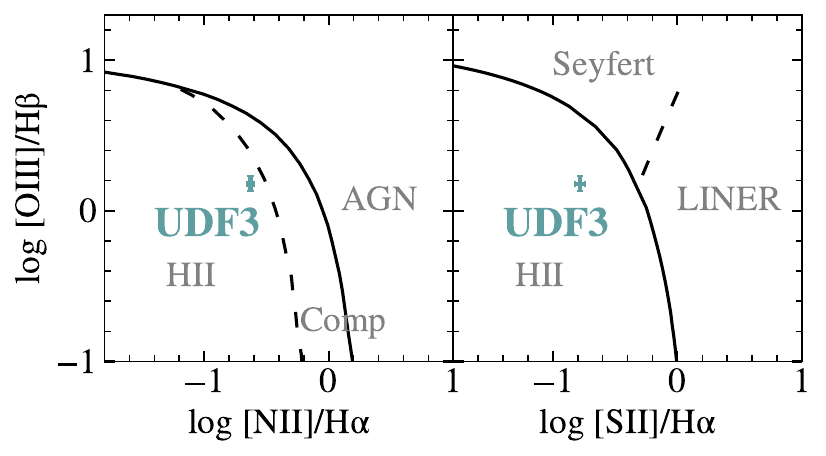}
    \caption{The BPT diagram, with the UDF3 plotted as blue filled circle and 1$\sigma$ error bar. The solid black lines show the maximum starburst boundaries from \citet{kewley2001theoretical}, while the dashed black lines in the left panel show the empirical maximum starburst line from \citet{kauffmann2003host}, and the right panel display the Seyfert-LINER separation lines from \citet{kewley2006host}.}
    \label{fig:BPT}
\end{figure}

The BPT diagram (\citealt{baldwin1981classification}) employs \nii/H$\alpha$ versus \oiii/H$\beta$ ratios to distinguish star-forming galaxies (lower ratios, UV-driven) from AGNs (higher ratios, accretion disk ionization). The ``star-forming sequence" emerges from this dichotomy. \citet{veilleux1987spectral} also proposed additional diagnostic diagrams (\sii/H$\alpha$ vs. \oiii/H$\beta$; [\ion{O}{1}] $\lambda$6300/H$\alpha$ vs. \oiii/H$\beta$) for classification. 

In this study, we use the [\ion{N}{2}]/H$\alpha$ versus [\ion{O}{3}]/H$\beta$ and [\ion{S}{2}]/H$\alpha$ versus [\ion{O}{3}]/H$\beta$ diagrams, with measured line ratios of  
log(\nii/H$\alpha$) $=-0.63\pm0.03$,  
log(\oiii/H$\beta$) $=0.18\pm0.05$,  
and 
log(\sii/H$\alpha$) $=-0.78\pm0.04$.
As shown in Figure~\ref{fig:BPT}, the ionized emission lines of UDF3 are well-explained by star formation. The demarcation lines in the BPT diagrams have been rigorously defined through observational and theoretical studies. The ``maximum starburst line" from \citet{kewley2001theoretical}, derived from stellar population synthesis and photoionization models, marks the upper limit of ionization achievable by young stars. Galaxies above this line are typically AGN-dominated. For composite systems, the empirical boundary by \citet{kauffmann2003host} separates star formation-dominated galaxies from those with mixed ionization sources. \citet{kewley2006host} added Seyfert-LINER separation lines.

At high redshift, the BPT classification boundaries shift towards the AGN region (e.g. \citealt{kewley2013cosmic}; \citealt{bian2020drives}; \citealt{garg2022bpt}), primarily due to lower metallicities and higher ionization parameters, which move the star-forming sequence to higher [\ion{O}{3}]/H$\beta$ and [\ion{N}{2}]/H$\alpha$ ratios. Therefore, this shift does not impact the classification of UDF3's ionized emission lines, which are dominated by \ion{H}{2} region-like activity.

\subsection{SED Fitting Suggests No Need for AGN to Explain the Infrared Emission}

\begin{figure*}[htbp]
\centering
\includegraphics[width=\textwidth]{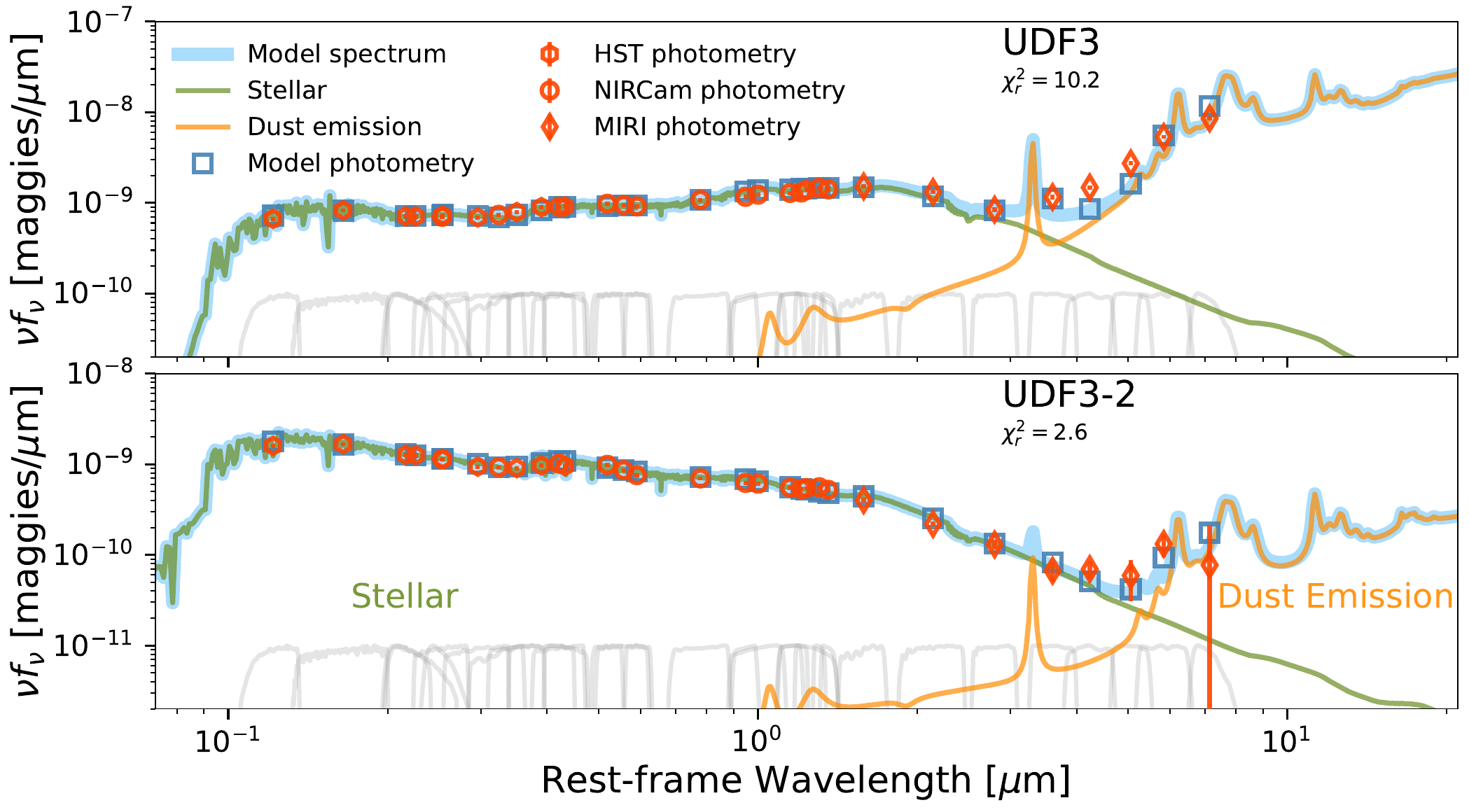} 
\caption{SED fitting of the two galaxies. The red hexagons, circles, and diamonds represent the photometric data from HST, NIRCam, and MIRI, respectively. The blue solid line and square markers denote the model spectrum and model photometry predicted by the SED fitting. The green solid line corresponds to the stellar component predicted by the \texttt{FSPS} model, while the orange solid line represents the fitted \citet{draine2007infrared} dust emission component. The AGN component is omitted in the plot due to its negligible contribution. The transmission curves of each filter are shown in gray line.}
\label{fig:SED}
\end{figure*}

Based on the SED fitting, as shown in Figure \ref{fig:SED} and Table \ref{tab:sed}, we find that the UV to mid-infrared photometry of UDF3 and UDF3-2 can be fully explained by their stellar components and dust emission heated by starlight, while the contribution from the AGN dust torus is negligible, with \( f_{\text{AGN}} < 10^{-6} \). The sSFR of both member galaxies exceed \( 10^{-8.5} \) yr$^{-1}$, indicating that HMXBs dominate the X-ray emission of this system.
Specifically, UDF3-2 is a star-forming galaxy located on the star-forming main sequence (SFMS) with \( M_\star =3.3^{+1.0}_{-0.6} \times 10^9 M_\odot \) and \( \text{SFR}= 34^{+6}_{-6}~M_\odot \, \text{yr}^{-1} \), exhibiting a relatively steady and continuous star formation history lasting $206^{+134}_{-66}$ Myr. In contrast, UDF3 is a dusty starburst galaxy with \( M_\star= 5.8^{+0.8}_{-0.5} \times 10^{9} M_\odot \) and \( \text{SFR} = 529_{-88}^{+64}~M_\odot \, \text{yr}^{-1} \), characterized by a short and intense star formation event lasting about 15 Myr. This timescale coincides with the peak of core-collapse supernova rates and HMXB formation efficiency \citep{shtykovskiy2007high}, resulting in a sufficiently large population of HMXBs and bright X-ray emission. 
Given the relatively short best-fit stellar age derived for UDF3, there may be concern about the presence of an underlying older stellar population not well captured by the delayed-$\tau$ SFH. To assess this, we perform an additional SED fitting using a more flexible non-parametric SFH with the ``LogM” prior introduced by \citet{leja2019measure}. We follow their Equation (2) to define a 7-bin time grid and rescale the time bins according to the cosmic age at $z = 2.544$. This non-parametric fit yields a stellar mass of $9.3_{-1.6}^{+6.0}\times10^{9}$ M$_\odot$ and an SFR of $585_{-91}^{+97}$~$M_\odot$~yr$^{-1}$. The resulting SFH closely matches that obtained with the delayed-$\tau$ model, suggesting that while an older stellar population may be present in UDF3, its contribution is minimal. This supports the interpretation that UDF3 is currently undergoing a recent burst of star formation that dominates its SED, consistent with the scenario in which its X-ray emission is powered by HMXBs.

UDF3 also exhibits high dust attenuation with \( A_{\text{V}} = 2.5\text{log}_{10}e\times \texttt{dust2} \approx 2.2 \) given by SED fitting and $A_{\text{V}} = 1.9 \pm 0.4$ given by Balmer decrement. The discrepancies in the model photometry, including an underestimation of the F1500W and F1800W points for UDF3 and overestimation of the F1280W, F1500W, and F2550W points for UDF3-2, are likely due to the incomplete treatment of PAH emission features in the 3--5 µm range within the \citet{draine2007infrared} dust emission model, as compared to more detailed models (e.g., \citealt{rieke2009determining}; \citealt{lyu2022agn}). We consider that mid-infrared photometry measurements at rest frame wavelengths longer than 2 µm, along with the modeling of dust emission in star-forming galaxies, are crucial for decomposing the dust emission and AGN components in SED fitting.

\subsection{X-ray can be naturally interpreted by HMXB host galaxy pair system}

\begin{figure}
    \centering
    \includegraphics[width=\linewidth]{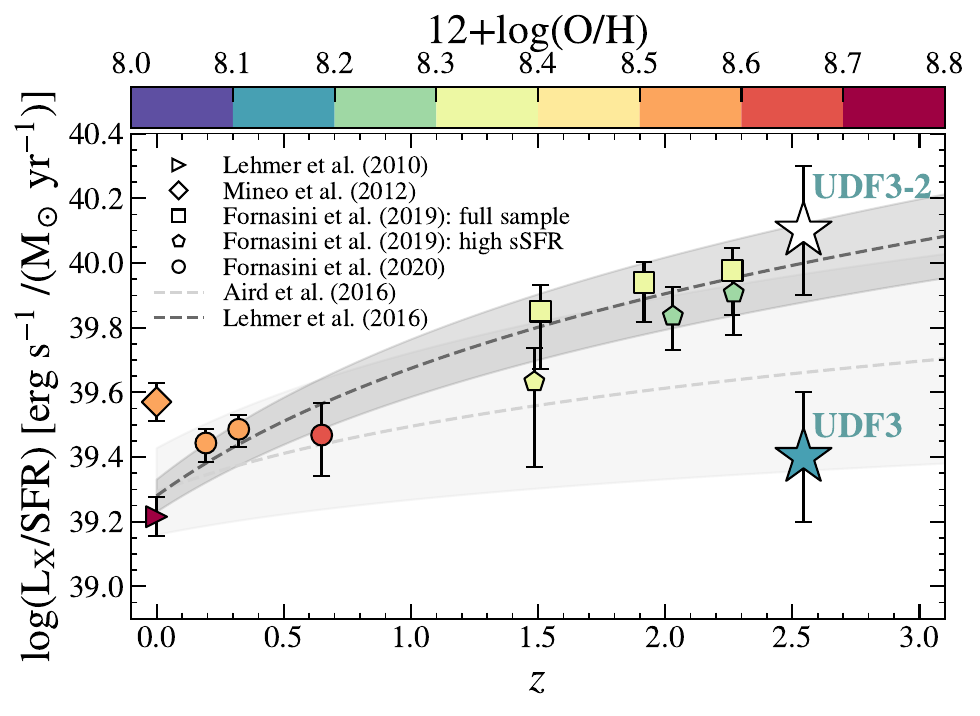} 
    \caption{The evolution of $L_X / \text{SFR}$ with redshift. The galaxy pair UDF3 and UDF3-2 are represented by star markers. Additionally, stacked $L_X / \text{SFR}$ values from previous studies in different redshift bins are marked. The metallicity, indicated by oxygen abundance, is shown in color. These include measurements at $z$=0 from \citet{lehmer2010chandra} and \citet{mineo2012x}, as well as from the full sample of \citet{fornasini2019mosdef}, and measurements from galaxies dominated by HMXBs ($\log\text{sSFR} > -8.8$). Measurements from \citet{fornasini2020connecting} in the hCOSMOS and zCOSMOS surveys are also displayed. The dark gray dashed line shows the redshift evolution of $L_X / \text{SFR}$ for HMXBs from \citet{lehmer2016evolution}: $L_X = \beta(1+z)^{\delta}\text{SFR}$. The light gray dashed line represents the relation measured by \citet{aird2016x}: $L_X = \beta(1+z)^{\delta}\text{SFR}^{\theta}$, normalized at SFR = 30 M$_\odot$yr$^{-1}$.}
    \label{fig:Lx_SFR}
\end{figure}

In this section, we present that the X-ray 
emission can be naturally interpreted by X-ray binary rather than AGN. Thus, UDF3 and UDF3-2 could be the highest-redshift X-ray binary galaxy pair with individual X-ray detections to date, at a redshift of $z=2.544$, located at the peak of cosmic star formation (Figure \ref{fig:Lx_SFR}).

The galaxy pair is likely undergoing a merger or interaction. During this process, UDF3 has entered a starburst phase, accompanied by rapid star formation. The short-wavelength side of the UDF3 image displays prominent clumpy structures (Figure \ref{fig:images} c, d, e), while UDF3-2 may exhibit an elongated morphology due to tidal forces. Previous studies have demonstrated that merger events typically trigger rapid star formation (e.g. \citealt{li2008interaction}; \citealt{patton2020interacting}; \citealt{bottrell2024illustristng}) and promote the growth of clumps (e.g. \citealt{calabro2019merger}; \citealt{nakazato2024merger}), and our observations are consistent with this theory.

Both galaxies exhibit sSFRs above $10^{-10}$ yr$^{-1}$, suggesting that their X-ray luminosities are likely dominated by HMXBs. 
Specifically, the UDF3-2, located on SFMS with 
log(sSFR) $=-8.0_{-0.2}^{+0.2}~\text{yr}^{-1}$ and log($L_X$/SFR) $=40.1^{+0.2}_{-0.2}$ erg s$^{-1}$/(M$_\odot$ yr$^{-1}$), is consistent with the trend of the $L_X$/SFR evolution with redshift observed in previous studies. However, due to the absence of emission-line spectral coverage, we are unable to determine its oxygen abundance. In contrast, UDF3 lies $\approx$ 0.5 dex below the trend with 
log($L_X$/SFR) $=39.4^{+0.2}_{-0.2}$ erg s$^{-1}$/(M$_\odot$ yr$^{-1}$). The reason is that UDF3 is currently undergoing a starburst phase triggered by galaxy interaction. It has a log(sSFR) $=-7.0_{-0.1}^{+0.1} ~\text{yr}^{-1}$ significantly higher than the typical values observed in the $L_X$/SFR-$z$ relation, where $-10<\log(\text{sSFR})<-8$. The oxygen abundance of UDF3 is 12 + log(O/H) = 8.13 $\pm$ 0.05. Although its average gas-phase metallicity is lower than that at lower redshifts as shown in Figure \ref{fig:Lx_SFR}, intense star formation and subsequent rapid supernova explosions within a short timescale may still have enhanced the metallicity of its stellar population. Under these conditions, although the high HMXB formation efficiency (discussed in Section 4.4) leads to enhanced X-ray emission in UDF3—accounting for its observable X-ray luminosity at high redshift—the elevated sSFR-driven stellar metal enrichment nevertheless suppresses Roche-lobe overflow efficiency, ultimately causing the deviation of $L_X$/SFR from the established trend.

The absorption of X-ray emission by dust and gas does not significantly affect our conclusions. Adopting a typical value of $\log(A_{\text{V}}/N_{\text{H}}) \approx -22$ (\citealt{guver2009relation}; \citealt{kaplan2010hi}) and estimating the intrinsic X-ray luminosity of the dusty galaxy UDF3 using the Portable, Interactive, Multi-Mission Simulator (PIMMS\footnote{\url{http://cxc.harvard.edu/toolkit/pimms.jsp}}), we find that the underestimation of $L_X$/SFR is less than 0.1 dex. 

SED modeling with \texttt{Prospector} may introduce systematic uncertainties in SFR estimates. \citet{haskell2024beware} showed that the delayed-$\tau$ SFH assumption underestimates the 100 Myr-averaged SFR by $\sim$0.11 dex. While this bias is mitigated during starbursts, uncertainties remain significantly underestimated. 
To cross-check our results, we derive the H$\alpha$-based SFR who has different systematics than SED-derived SFR. 
The dust extinction-corrected H$\alpha$ luminosity using the reddening curve from \citet{kriek2013dust} is $L_{\text{H}\alpha} = (1.9 \pm 0.9)\times10^{-17}$ erg s$^{-1}$, yielding SFR(H$\alpha$) = $153\pm69$ M$_\odot$ yr$^{-1}$ under the calibration from \citet{kennicutt1998star}.

In previous studies, X-ray luminosity has often been used as evidence for the presence of AGN. For example, \citet{luo2016chandra} classified sources with \( L_{X,0.5-7 \, \text{keV}} > 3\times10^{42} \, \text{erg s$^{-1}$} \) as AGN, while \citet{fornasini2019mosdef} suggested that sources with \( L_{X,2-10 \, \text{keV}} > 10^{41.5} \, \text{erg s$^{-1}$} \) could be influenced by AGN. However, our multi-band analysis indicates that previous studies may have underestimated the contribution of X-ray emission from star-forming galaxies to the overall cosmic X-ray luminosity, particularly the role played by the X-ray binary populations in these galaxies.

\section{Conclusion} \label{sec:conclusion}

In this work, we use photometric data from the JADES survey, spectroscopic data from the DJA database, and Chandra 7 Ms deep data to identify a galaxy pair at \( z = 2.544 \) in the GOODS-S field.

The BPT diagram of UDF3 shows that the ionized emission lines, particularly [\ion{N}{2}]/H$\alpha$ versus [\ion{O}{3}]/H$\beta$ and [\ion{S}{2}]/H$\alpha$ versus [\ion{O}{3}]/H$\beta$, are dominated by star formation in \ion{H}{2} regions. SED decomposition reveals that both galaxies are primarily explained by stellar and dust components, with an AGN fraction of \( f_{\text{AGN}} < 10^{-6} \). UDF3 is a dusty starburst galaxy, with dust heated by young high-temperature O/B stars, emitting bright infrared radiation. UDF3-2 is a star-forming galaxy on the main sequence. The galaxies likely interact gravitationally, with UDF3 exhibiting observable clumpy structures and UDF3-2 showing an elongated morphology due to tidal forces.

This galaxy pair could be  
the highest-redshift star-forming pair with 
X-ray detected in individual member galaxy.  

After careful examination, we conclude that X-ray binary rather than 
AGN dominate the X-ray emission. 
The sSFR of both member galaxies exceed \( 10^{-8.5} \) yr$^{-1}$, with UDF3 particularly undergoing the peak timescale for HMXB formation efficiency, implying HMXBs to 
play a dominant role in their X-ray appearance. 
We also compare them to the $L_X$/SFR evolution trend with 
redshift in the previous works. In the context of decreasing cosmic metallicity, UDF3, 
with a metallicity of 12 + log(O/H) = 8.13 $\pm$ 0.05 , deviates from the star-forming main sequence, 
leading to a lower Lx/SFR. In contrast, UDF3-2 aligns more closely with the trend.

The discovery of this galaxy pair suggests that 
X-ray luminosity may not completely come from AGN activities. 
Star formation event, especially HMXB activities may be 
considered when estimating the overall X-ray luminosity of the Universe. 
With the release of more MIRI data, we can more precisely identify XRB host galaxy candidates at \( z > 2 \), combining this with deep X-ray surveys to increase the statistical sample. This will deepen our understanding of the evolutionary history of XRB populations associated with star formation activities across the cosmic time.

\section{Acknowledgments}

We would like to thank Prof. Dandan Xu, Prof. Bin Luo and Qinnan Zhu for helpful discussions. 
We acknowledge support from the National Key R\&D Program of China (grant no. 2023YFA1605600) and Tsinghua University Initiative Scientific Research Program (No. 20223080023).
We acknowledge use of JWST and HST data from the JWST Advanced Deep Extragalactic Survey (JADES) and the Systematic MIRI Legacy Extragalactic Survey (SMILES), provided by their respective teams via the Mikulski Archive for Space Telescopes (MAST) at the Space Telescope Science Institute.
The specific observations analyzed can be accessed via \dataset[doi: 10.17909/8tdj-8n28]{https://archive.stsci.edu/doi/resolve/resolve.html?doi=10.17909/8tdj-8n28} and \dataset[doi: 10.17909/jmxm-1695]{https://archive.stsci.edu/doi/resolve/resolve.html?doi=10.17909/jmxm-1695}.
Some of the data products presented herein were retrieved from the Dawn JWST Archive (DJA). DJA is an initiative of the Cosmic Dawn Center (DAWN), which is funded by the Danish National Research Foundation under grant DNRF140. 
This paper uses X‑ray data products from the 7~Ms Chandra Deep Field‑South survey, as provided in the publicly available catalog hosted on VizieR (with \dataset[DOI: 10.26093/cds/vizier.22280002]{https://doi.org/10.26093/cds/vizier.22280002}).



%

\vspace{5mm}
\facilities{HST (ACS, WFC3), JWST (NIRCam, MIRI, NIRSpec), VLT (VIMOS), Chandra (ACIS).}


\software{Astropy \citep{robitaille2013astropy,price2018astropy,price2022astropy}, Prospector \citep{johnson2021stellar}, FSPS \citep{conroy2010propagation}, dynesty \citep{speagle2020dynesty}, Matplotlib \citep{hunter2007matplotlib}, NumPy \citep{harris2020array}, SciPy \citep{virtanen2020scipy}
}



\appendix

In the table \ref{tab:info}, we list the IDs of the two galaxies from the JADES and SMILES photometric surveys, along with their counterparts in the CDFS 7 Ms source catalog and the NIRSpec and VANDELS spectroscopic surveys. Notably, the NIRSpec spectra of UDF3 were obtained under JWST Proposal 1180, with a source ID of 14279. Also, we list all the photometry results before corrected by emission lines. In the tabel \ref{tab:fluxes}, we show the emission-line fluxes of UDF3 detected by the NIRSpec grating spectra described in the Section \ref{sec:Emission-line Fluxes}.

\begin{table*}[htbp]
\centering
\caption{Basic Information and Photometry of the Galaxy Pair}
\begin{tabular}{cccccccc}
\toprule
\toprule
Galaxy ID & JADES ID & SMILES ID & XID & NIRSpec ID & VANDELS ID & RA$_{\text{JADES}}$ & DEC$_{\text{JADES}}$ \\
\midrule
UDF3   &  209117  &  1271  & 718  & 1180-14279 & - & 53.160609 & 	-27.776249\\
UDF3-2 &  209116  &  1269  &  -  &  - &  CDFS015798 & 53.160157 & 	-27.776414\\
\midrule
acs\_wfc\_f435w & acs\_wfc\_f606w & acs\_wfc\_f775w & acs\_wfc\_f814w & acs\_wfc\_f850lp & wfc3\_ir\_f105w & wfc3\_ir\_f125w & wfc3\_ir\_f140w \\
\midrule
25.23 $\pm$ 0.02 & 24.66 $\pm$ 0.01 & 24.54 $\pm$ 0.01 & 24.49 $\pm$ 0.05 & 24.34 $\pm$ 0.02 & 24.22 $\pm$ 0.02 & 23.89 $\pm$ 0.02 & 23.64 $\pm$ 0.01 \\
24.28 $\pm$ 0.01 & 23.90 $\pm$ 0.01 & 23.90 $\pm$ 0.01 & 23.87 $\pm$ 0.03 & 23.85 $\pm$ 0.01 & 23.89 $\pm$ 0.02 & 23.74 $\pm$ 0.02 & 23.56 $\pm$ 0.02\\
\midrule
wfc3\_ir\_f160w & nircam\_f090w & nircam\_f115w & nircam\_f150w & nircam\_f182m & nircam\_f200w & nircam\_f210m & nircam\_f277w \\
\midrule
23.51 $\pm$ 0.02 & 24.37 $\pm$ 0.02 & 24.05 $\pm$ 0.01 & 23.52 $\pm$ 0.01 & 23.20 $\pm$ 0.01 & 23.18 $\pm$ 0.01 & 23.17 $\pm$ 0.01 & 22.69 $\pm$ 0.00\\
23.46 $\pm$ 0.02 & 23.85 $\pm$ 0.01 & 23.81 $\pm$ 0.01 & 23.42 $\pm$ 0.01 & 23.23 $\pm$ 0.01 & 23.30 $\pm$ 0.01 & 23.37 $\pm$ 0.01 & 23.16 $\pm$ 0.01\\
\midrule
nircam\_f335m & nircam\_f356w & nircam\_f410m & nircam\_f430m & nircam\_f444w & nircam\_f460m & nircam\_f480m & miri\_f560w \\
\midrule
22.36 $\pm$ 0.00 & 22.25 $\pm$ 0.00 & 22.07 $\pm$ 0.00 & 22.01 $\pm$ 0.00 & 21.91 $\pm$ 0.00 & 21.77 $\pm$ 0.00 & 21.80 $\pm$ 0.00 & 21.55 $\pm$ 0.02 \\
23.07 $\pm$ 0.01 & 23.03 $\pm$ 0.00 & 22.99 $\pm$ 0.01 & 22.98 $\pm$ 0.01 & 22.93 $\pm$ 0.00 & 22.84 $\pm$ 0.01 & 22.87 $\pm$ 0.01 & 22.99 $\pm$ 0.04 \\
\midrule
miri\_f770w & miri\_f1000w & miri\_f1280w & miri\_f1500w & miri\_f1800w & miri\_f2100w & miri\_f2550w &  \\
\midrule
21.37 $\pm$ 0.02 & 21.57 $\pm$ 0.04 & 20.95 $\pm$ 0.02 & 20.51 $\pm$ 0.02 & 19.65 $\pm$ 0.02 & 18.76 $\pm$ 0.01 & 18.04 $\pm$ 0.02 & \\
23.32 $\pm$ 0.05 & 23.60 $\pm$ 0.12 & 24.05 $\pm$ 0.24 & 23.83 $\pm$ 0.23 & 23.81 $\pm$ 0.52 & 22.78 $\pm$ 0.30 & 23.15 $\pm$ 1.92 & \\
\bottomrule
\end{tabular}
\tablecomments{The units of the photometric measurements and uncertainties are in AB magnitude. For both surveys, we adopt PSF-matched Kron fluxes based on JADES v0.9.0 catalog and SMILES v0.5 catalog.}
\label{tab:info}
\end{table*}

\begin{table}[h!]
\centering
\caption{Emission-line Fluxes of UDF3}
\begin{tabular}{cccc}
\toprule
\toprule
Emission Line & Flux ($10^{-18} \, \text{erg} \, \text{s}^{-1} \, \text{cm}^{-2}$) & Emission Line & Flux ($10^{-18} \, \text{erg} \, \text{s}^{-1} \, \text{cm}^{-2}$) \\
\midrule
H$\beta$ &  8.10~$\pm$~0.84 & \text{[\ion{S}{2}]} $\lambda6731$ & 3.35~$\pm$~0.45\\
\text{[\ion{O}{3}]} $\lambda4959$ & 4.44~$\pm$~0.68 & \text{[\ion{S}{3}]} $\lambda9069$ &  3.12~$\pm$~0.43\\
\text{[\ion{O}{3}]} $\lambda5007$ & 12.30~$\pm$~0.83 & Pa$\delta$ & 1.22~$\pm$~0.27\\
\text{[\ion{N}{2}]} $\lambda6548$ & 3.75~$\pm$~0.49 & \text{\ion{He}{1}} $\lambda10830$ & 7.07~$\pm$~0.43\\
H$\alpha$ &  39.97~$\pm$~0.95 & Pa$\gamma$ & 2.08~$\pm$~0.28\\
\text{[\ion{N}{2}]} $\lambda6584$ & 9.41~$\pm$~0.58 & \text{\ion{Fe}{2}} $\lambda12570$ & 1.16~$\pm$~0.39\\
\text{[\ion{S}{2}]} $\lambda6716$ & 3.26~$\pm$~0.46 & Pa$\beta$ & 6.04~$\pm$~0.58\\
\bottomrule
\end{tabular}
\label{tab:fluxes}
\end{table}


\bibliography{sample631}{}

\begin{thebibliography}{}
\expandafter\ifx\csname natexlab\endcsname\relax\def\natexlab#1{#1}\fi
\providecommand{\url}[1]{\href{#1}{#1}}
\providecommand{\dodoi}[1]{doi:~\href{http://doi.org/#1}{\nolinkurl{#1}}}
\providecommand{\doeprint}[1]{\href{http://ascl.net/#1}{\nolinkurl{http://ascl.net/#1}}}
\providecommand{\doarXiv}[1]{\href{https://arxiv.org/abs/#1}{\nolinkurl{https://arxiv.org/abs/#1}}}

\bibitem[{Ade {et~al.}(2016)Ade, Aghanim, Arnaud, Ashdown, Aumont, Baccigalupi, Banday, Barreiro, Bartlett, Bartolo, {et~al.}}]{ade2016planck}
Ade, P.~A., Aghanim, N., Arnaud, M., {et~al.} 2016, Astronomy \& Astrophysics, 594, A13

\bibitem[{Aird {et~al.}(2016)Aird, Coil, \& Georgakakis}]{aird2016x}
Aird, J., Coil, A., \& Georgakakis, A. 2016, Monthly Notices of the Royal Astronomical Society, 465, 3390

\bibitem[{Alberts {et~al.}(2024)Alberts, Lyu, Shivaei, Rieke, Perez-Gonzalez, Bonventura, Zhu, Helton, Ji, Morrison, {et~al.}}]{alberts2024smiles}
Alberts, S., Lyu, J., Shivaei, I., {et~al.} 2024, arXiv preprint arXiv:2405.15972

\bibitem[{Antoniou \& Zezas(2016)}]{antoniou2016star}
Antoniou, V., \& Zezas, A. 2016, Monthly Notices of the Royal Astronomical Society, 459, 528

\bibitem[{Artale {et~al.}(2015)Artale, Tissera, \& Pellizza}]{artale2015stellar}
Artale, M.~C., Tissera, P.~B., \& Pellizza, L. 2015, Monthly Notices of the Royal Astronomical Society, 448, 3071

\bibitem[{Astropy~Collaboration {et~al.}(2018)Astropy~Collaboration, Sip{\H{o}}cz, G{\"u}nther, Lim, Crawford, Conseil, Shupe, Craig, Dencheva, Ginsburg, {et~al.}}]{price2018astropy}
Astropy~Collaboration, Price-Whelan, A.~M., Sip{\H{o}}cz, B., G{\"u}nther, H., {et~al.} 2018, The Astronomical Journal, 156, 123

\bibitem[{Astropy~Collaboration {et~al.}(2022)Astropy~Collaboration, Lim, Earl, Starkman, Bradley, Shupe, Patil, Corrales, Brasseur, N{\"o}the, {et~al.}}]{price2022astropy}
Astropy~Collaboration, Price-Whelan, A.~M., Lim, P.~L., Earl, N., {et~al.} 2022, The Astrophysical Journal, 935, 167

\bibitem[{Astropy~Collaboration {et~al.}(2013)Astropy~Collaboration, Tollerud, Greenfield, Droettboom, Bray, Aldcroft, Davis, Ginsburg, Price-Whelan, Kerzendorf, {et~al.}}]{robitaille2013astropy}
Astropy~Collaboration, Robitaille, T.~P., Tollerud, E.~J., Greenfield, P., {et~al.} 2013, Astronomy \& Astrophysics, 558, A33

\bibitem[{Bacon {et~al.}(2017)Bacon, Conseil, Mary, Brinchmann, Shepherd, Akhlaghi, Weilbacher, Piqueras, Wisotzki, Lagattuta, {et~al.}}]{bacon2017muse}
Bacon, R., Conseil, S., Mary, D., {et~al.} 2017, Astronomy \& Astrophysics, 608, A1

\bibitem[{Baldwin {et~al.}(1981)Baldwin, Phillips, \& Terlevich}]{baldwin1981classification}
Baldwin, J.~A., Phillips, M.~M., \& Terlevich, R. 1981, Publications of the Astronomical Society of the Pacific, 93, 5

\bibitem[{Basu-Zych {et~al.}(2012)Basu-Zych, Lehmer, Hornschemeier, Bouwens, Fragos, Oesch, Belczynski, Brandt, Kalogera, Luo, {et~al.}}]{basu2012x}
Basu-Zych, A.~R., Lehmer, B.~D., Hornschemeier, A.~E., {et~al.} 2012, The Astrophysical Journal, 762, 45

\bibitem[{Belczynski {et~al.}(2004)Belczynski, Kalogera, Zezas, \& Fabbiano}]{belczynski2004x}
Belczynski, K., Kalogera, V., Zezas, A., \& Fabbiano, G. 2004, The Astrophysical Journal, 601, L147

\bibitem[{Bian {et~al.}(2020)Bian, Kewley, Groves, \& Dopita}]{bian2020drives}
Bian, F., Kewley, L.~J., Groves, B., \& Dopita, M.~A. 2020, Monthly Notices of the Royal Astronomical Society, 493, 580

\bibitem[{Bodaghee {et~al.}(2011)Bodaghee, Tomsick, Rodriguez, \& James}]{bodaghee2011clustering}
Bodaghee, A., Tomsick, J.~A., Rodriguez, J., \& James, J.~B. 2011, The Astrophysical Journal, 744, 108

\bibitem[{Bottrell {et~al.}(2024)Bottrell, Yesuf, Popping, Omori, Tang, Ding, Pillepich, Nelson, Eisert, Gao, {et~al.}}]{bottrell2024illustristng}
Bottrell, C., Yesuf, H.~M., Popping, G., {et~al.} 2024, Monthly Notices of the Royal Astronomical Society, 527, 6506

\bibitem[{Brorby {et~al.}(2016)Brorby, Kaaret, Prestwich, \& Mirabel}]{brorby2016enhanced}
Brorby, M., Kaaret, P., Prestwich, A., \& Mirabel, I.~F. 2016, Monthly Notices of the Royal Astronomical Society, 457, 4081

\bibitem[{Buat {et~al.}(2019)Buat, Ciesla, Boquien, Ma{\l}ek, \& Burgarella}]{buat2019cold}
Buat, V., Ciesla, L., Boquien, M., Ma{\l}ek, K., \& Burgarella, D. 2019, Astronomy \& Astrophysics, 632, A79

\bibitem[{Calabr{\`o} {et~al.}(2019)Calabr{\`o}, Daddi, Fensch, Bournaud, Cibinel, Puglisi, Jin, Delvecchio, \& D’eugenio}]{calabro2019merger}
Calabr{\`o}, A., Daddi, E., Fensch, J., {et~al.} 2019, Astronomy \& Astrophysics, 632, A98

\bibitem[{Conroy \& Gunn(2010)}]{conroy2010propagation}
Conroy, C., \& Gunn, J.~E. 2010, The Astrophysical Journal, 712, 833

\bibitem[{Conroy {et~al.}(2009)Conroy, Gunn, \& White}]{conroy2009propagation}
Conroy, C., Gunn, J.~E., \& White, M. 2009, The Astrophysical Journal, 699, 486

\bibitem[{Cowie {et~al.}(2012)Cowie, Barger, \& Hasinger}]{cowie2012faintest}
Cowie, L., Barger, A., \& Hasinger, G. 2012, The Astrophysical Journal, 748, 50

\bibitem[{Cowie {et~al.}(2018)Cowie, Gonzalez-Lopez, Barger, Bauer, Hsu, \& Wang}]{cowie2018submillimeter}
Cowie, L.~L., Gonzalez-Lopez, J., Barger, A.~J., {et~al.} 2018, The Astrophysical Journal, 865, 106

\bibitem[{Das {et~al.}(2017)Das, Mesinger, Pallottini, Ferrara, \& Wise}]{das2017high}
Das, A., Mesinger, A., Pallottini, A., Ferrara, A., \& Wise, J.~H. 2017, Monthly Notices of the Royal Astronomical Society, 469, 1166

\bibitem[{Dijkstra {et~al.}(2012)Dijkstra, Gilfanov, Loeb, \& Sunyaev}]{dijkstra2012constraints}
Dijkstra, M., Gilfanov, M., Loeb, A., \& Sunyaev, R. 2012, Monthly Notices of the Royal Astronomical Society, 421, 213

\bibitem[{Draine \& Li(2007)}]{draine2007infrared}
Draine, B.~T., \& Li, A. 2007, The Astrophysical Journal, 657, 810

\bibitem[{Dunlop {et~al.}(2017)Dunlop, McLure, Biggs, Geach, Micha{\l}owski, Ivison, Rujopakarn, van Kampen, Kirkpatrick, Pope, {et~al.}}]{dunlop2017deep}
Dunlop, J., McLure, R.~J., Biggs, A.~D., {et~al.} 2017, Monthly Notices of the Royal Astronomical Society, 466, 861

\bibitem[{Eisenstein {et~al.}(2023)Eisenstein, Willott, Alberts, Arribas, Bonaventura, Bunker, Cameron, Carniani, Charlot, Curtis-Lake, {et~al.}}]{eisenstein2023overview}
Eisenstein, D.~J., Willott, C., Alberts, S., {et~al.} 2023, arXiv preprint arXiv:2306.02465

\bibitem[{Ferrara {et~al.}(2023)Ferrara, Pallottini, \& Dayal}]{ferrara2023stunning}
Ferrara, A., Pallottini, A., \& Dayal, P. 2023, Monthly Notices of the Royal Astronomical Society, 522, 3986

\bibitem[{Fornasini {et~al.}(2020)Fornasini, Civano, \& Suh}]{fornasini2020connecting}
Fornasini, F.~M., Civano, F., \& Suh, H. 2020, Monthly Notices of the Royal Astronomical Society, 495, 771

\bibitem[{Fornasini {et~al.}(2019)Fornasini, Kriek, Sanders, Shivaei, Civano, Reddy, Shapley, Coil, Mobasher, Siana, {et~al.}}]{fornasini2019mosdef}
Fornasini, F.~M., Kriek, M., Sanders, R.~L., {et~al.} 2019, The Astrophysical Journal, 885, 65

\bibitem[{Fragos {et~al.}(2013)Fragos, Lehmer, Tremmel, Tzanavaris, Basu-Zych, Belczynski, Hornschemeier, Jenkins, Kalogera, Ptak, {et~al.}}]{fragos2013x}
Fragos, T., Lehmer, B., Tremmel, M., {et~al.} 2013, The Astrophysical Journal, 764, 41

\bibitem[{Franco {et~al.}(2018)Franco, Elbaz, B{\'e}thermin, Magnelli, Schreiber, Ciesla, Dickinson, Nagar, Silverman, Daddi, {et~al.}}]{franco2018goods}
Franco, M., Elbaz, D., B{\'e}thermin, M., {et~al.} 2018, Astronomy \& Astrophysics, 620, A152

\bibitem[{Gallo {et~al.}(2005)Gallo, Fender, Kaiser, Russell, Morganti, Oosterloo, \& Heinz}]{gallo2005dark}
Gallo, E., Fender, R., Kaiser, C., {et~al.} 2005, Nature, 436, 819

\bibitem[{Garg {et~al.}(2022)Garg, Narayanan, Byler, Sanders, Shapley, Strom, Dav{\'e}, Hirschmann, Lovell, Otter, {et~al.}}]{garg2022bpt}
Garg, P., Narayanan, D., Byler, N., {et~al.} 2022, The Astrophysical Journal, 926, 80

\bibitem[{Garilli {et~al.}(2021)Garilli, McLure, Pentericci, Franzetti, Gargiulo, Carnall, Cucciati, Iovino, Amorin, Bolzonella, {et~al.}}]{garilli2021vandels}
Garilli, B., McLure, R., Pentericci, L., {et~al.} 2021, Astronomy \& Astrophysics, 647, A150

\bibitem[{Gilfanov(2004)}]{gilfanov2004low}
Gilfanov, M. 2004, Monthly Notices of the Royal Astronomical Society, 349, 146

\bibitem[{Gilfanov {et~al.}(2023)Gilfanov, Fabbiano, Lehmer, \& Zezas}]{gilfanov2023x}
Gilfanov, M., Fabbiano, G., Lehmer, B., \& Zezas, A. 2023, in Handbook of X-ray and Gamma-ray Astrophysics (Springer), 1--38

\bibitem[{Grimm {et~al.}(2003)Grimm, Gilfanov, \& Sunyaev}]{grimm2003high}
Grimm, H.-J., Gilfanov, M., \& Sunyaev, R. 2003, Monthly Notices of the Royal Astronomical Society, 339, 793

\bibitem[{G{\"u}ver \& {\"O}zel(2009)}]{guver2009relation}
G{\"u}ver, T., \& {\"O}zel, F. 2009, Monthly Notices of the Royal Astronomical Society, 400, 2050

\bibitem[{Harris {et~al.}(2020)Harris, Millman, Van Der~Walt, Gommers, Virtanen, Cournapeau, Wieser, Taylor, Berg, Smith, {et~al.}}]{harris2020array}
Harris, C.~R., Millman, K.~J., Van Der~Walt, S.~J., {et~al.} 2020, Nature, 585, 357

\bibitem[{Haskell {et~al.}(2024)Haskell, Das, Smith, Cochrane, Hayward, \& Angl{\'e}s-Alc{\'a}zar}]{haskell2024beware}
Haskell, P., Das, S., Smith, D., {et~al.} 2024, Monthly Notices of the Royal Astronomical Society: Letters, 530, L7

\bibitem[{Heintz {et~al.}(2024)}]{heintz2024jwst}
Heintz, K., {et~al.} 2024, Preprint at https://arxiv. org/abs/2404.02211

\bibitem[{Hunter(2007)}]{hunter2007matplotlib}
Hunter, J.~D. 2007, Computing in science \& engineering, 9, 90

\bibitem[{Iben~Jr {et~al.}(1995)Iben~Jr, Tutukov, \& Yungelson}]{iben1995model}
Iben~Jr, I., Tutukov, A.~V., \& Yungelson, L.~R. 1995, Astrophysical Journal Supplement v. 100, p. 217, 100, 217

\bibitem[{Johnson {et~al.}(2021)Johnson, Leja, Conroy, \& Speagle}]{johnson2021stellar}
Johnson, B.~D., Leja, J., Conroy, C., \& Speagle, J.~S. 2021, The Astrophysical Journal Supplement Series, 254, 22

\bibitem[{Kaplan {et~al.}(2010)Kaplan, Prochaska, Herbert-Fort, Ellison, \& Dessauges-Zavadsky}]{kaplan2010hi}
Kaplan, K.~F., Prochaska, J.~X., Herbert-Fort, S., Ellison, S.~L., \& Dessauges-Zavadsky, M. 2010, Publications of the Astronomical Society of the Pacific, 122, 619

\bibitem[{Kauffmann {et~al.}(2003)Kauffmann, Heckman, Tremonti, Brinchmann, Charlot, White, Ridgway, Brinkmann, Fukugita, Hall, {et~al.}}]{kauffmann2003host}
Kauffmann, G., Heckman, T.~M., Tremonti, C., {et~al.} 2003, Monthly Notices of the Royal Astronomical Society, 346, 1055

\bibitem[{Kennicutt~Jr(1998)}]{kennicutt1998star}
Kennicutt~Jr, R.~C. 1998, Annual Review of Astronomy and Astrophysics, 36, 189

\bibitem[{Kewley {et~al.}(2001)Kewley, Dopita, Sutherland, Heisler, \& Trevena}]{kewley2001theoretical}
Kewley, L.~J., Dopita, M.~A., Sutherland, R., Heisler, C., \& Trevena, J. 2001, The Astrophysical Journal, 556, 121

\bibitem[{Kewley {et~al.}(2006)Kewley, Groves, Kauffmann, \& Heckman}]{kewley2006host}
Kewley, L.~J., Groves, B., Kauffmann, G., \& Heckman, T. 2006, Monthly Notices of the Royal Astronomical Society, 372, 961

\bibitem[{Kewley {et~al.}(2013)Kewley, Maier, Yabe, Ohta, Akiyama, Dopita, \& Yuan}]{kewley2013cosmic}
Kewley, L.~J., Maier, C., Yabe, K., {et~al.} 2013, The Astrophysical Journal Letters, 774, L10

\bibitem[{Kriek \& Conroy(2013)}]{kriek2013dust}
Kriek, M., \& Conroy, C. 2013, The Astrophysical Journal Letters, 775, L16

\bibitem[{Kroupa(2001)}]{kroupa2001variation}
Kroupa, P. 2001, Monthly Notices of the Royal Astronomical Society, 322, 231

\bibitem[{Lehmer {et~al.}(2010)Lehmer, Alexander, Bauer, Brandt, Goulding, Jenkins, Ptak, \& Roberts}]{lehmer2010chandra}
Lehmer, B., Alexander, D., Bauer, F., {et~al.} 2010, The Astrophysical Journal, 724, 559

\bibitem[{Lehmer {et~al.}(2016)Lehmer, Basu-Zych, Mineo, Brandt, Eufrasio, Fragos, Hornschemeier, Luo, Xue, Bauer, {et~al.}}]{lehmer2016evolution}
Lehmer, B., Basu-Zych, A., Mineo, S., {et~al.} 2016, The Astrophysical Journal, 825, 7

\bibitem[{Lehmer {et~al.}(2019)Lehmer, Eufrasio, Tzanavaris, Basu-Zych, Fragos, Prestwich, Yukita, Zezas, Hornschemeier, \& Ptak}]{lehmer2019x}
Lehmer, B.~D., Eufrasio, R.~T., Tzanavaris, P., {et~al.} 2019, The Astrophysical Journal Supplement Series, 243, 3

\bibitem[{Lehmer {et~al.}(2024)Lehmer, Monson, Eufrasio, Amiri, Doore, Basu-Zych, Garofali, Oskinova, Andrews, Antoniou, {et~al.}}]{lehmer2024empirical}
Lehmer, B.~D., Monson, E.~B., Eufrasio, R.~T., {et~al.} 2024, The Astrophysical Journal, 977, 189

\bibitem[{Leja {et~al.}(2019)Leja, Carnall, Johnson, Conroy, \& Speagle}]{leja2019measure}
Leja, J., Carnall, A.~C., Johnson, B.~D., Conroy, C., \& Speagle, J.~S. 2019, The Astrophysical Journal, 876, 3

\bibitem[{Lewin {et~al.}(1997)Lewin, van~den Heuvel, \& van Paradijs}]{lewin1997x}
Lewin, W.~H., van~den Heuvel, E.~P., \& van Paradijs, J. 1997, X-ray Binaries, Vol.~26 (Cambridge University Press)

\bibitem[{Li {et~al.}(2008)Li, Kauffmann, Heckman, Jing, \& White}]{li2008interaction}
Li, C., Kauffmann, G., Heckman, T.~M., Jing, Y., \& White, S.~D. 2008, Monthly Notices of the Royal Astronomical Society, 385, 1903

\bibitem[{Linden {et~al.}(2010)Linden, Kalogera, Sepinsky, Prestwich, Zezas, \& Gallagher}]{linden2010effect}
Linden, T., Kalogera, V., Sepinsky, J., {et~al.} 2010, The Astrophysical Journal, 725, 1984

\bibitem[{Llerena {et~al.}(2024)Llerena, Pentericci, Napolitano, Mascia, Amor{\'\i}n, Calabr{\`o}, Castellano, Cleri, Giavalisco, Grogin, {et~al.}}]{llerena2024ionizing}
Llerena, M., Pentericci, L., Napolitano, L., {et~al.} 2024, arXiv preprint arXiv:2412.01358

\bibitem[{Luo {et~al.}(2016)Luo, Brandt, Xue, Lehmer, Alexander, Bauer, Vito, Yang, Basu-Zych, Comastri, {et~al.}}]{luo2016chandra}
Luo, B., Brandt, W., Xue, Y., {et~al.} 2016, The Astrophysical Journal Supplement Series, 228, 2

\bibitem[{Lyu {et~al.}(2022)Lyu, Alberts, Rieke, \& Rujopakarn}]{lyu2022agn}
Lyu, J., Alberts, S., Rieke, G.~H., \& Rujopakarn, W. 2022, The Astrophysical Journal, 941, 191

\bibitem[{Lyu {et~al.}(2024)Lyu, Alberts, Rieke, Shivaei, P{\'e}rez-Gonz{\'a}lez, Sun, Hainline, Baum, Bonaventura, Bunker, {et~al.}}]{lyu2024active}
Lyu, J., Alberts, S., Rieke, G.~H., {et~al.} 2024, The Astrophysical Journal, 966, 229

\bibitem[{Madau \& Fragos(2017)}]{madau2017radiation}
Madau, P., \& Fragos, T. 2017, The Astrophysical Journal, 840, 39

\bibitem[{Mineo {et~al.}(2012)Mineo, Gilfanov, \& Sunyaev}]{mineo2012x}
Mineo, S., Gilfanov, M., \& Sunyaev, R. 2012, Monthly Notices of the Royal Astronomical Society, 419, 2095

\bibitem[{Misra {et~al.}(2023)Misra, Kovlakas, Fragos, Lazzarini, Bavera, Lehmer, Zezas, Zapartas, Xing, Andrews, {et~al.}}]{misra2023x}
Misra, D., Kovlakas, K., Fragos, T., {et~al.} 2023, Astronomy \& Astrophysics, 672, A99

\bibitem[{Nakazato {et~al.}(2024)Nakazato, Ceverino, \& Yoshida}]{nakazato2024merger}
Nakazato, Y., Ceverino, D., \& Yoshida, N. 2024, arXiv preprint arXiv:2402.08911

\bibitem[{Nenkova {et~al.}(2008)Nenkova, Sirocky, Nikutta, Ivezi{\'c}, \& Elitzur}]{nenkova2008agn}
Nenkova, M., Sirocky, M.~M., Nikutta, R., Ivezi{\'c}, {\v{Z}}., \& Elitzur, M. 2008, The Astrophysical Journal, 685, 160

\bibitem[{Pantoni {et~al.}(2021{\natexlab{a}})Pantoni, Lapi, Massardi, Donevski, Bressan, Silva, Pozzi, Vignali, Talia, Cimatti, {et~al.}}]{pantoni2021unveiling}
Pantoni, L., Lapi, A., Massardi, M., {et~al.} 2021{\natexlab{a}}, Monthly Notices of the Royal Astronomical Society, 504, 928

\bibitem[{Pantoni {et~al.}(2021{\natexlab{b}})Pantoni, Massardi, Lapi, Donevski, D’Amato, Giulietti, Pozzi, Talia, Vignali, Cimatti, {et~al.}}]{pantoni2021alma}
Pantoni, L., Massardi, M., Lapi, A., {et~al.} 2021{\natexlab{b}}, Monthly Notices of the Royal Astronomical Society, 507, 3998

\bibitem[{Patton {et~al.}(2020)Patton, Wilson, Metrow, Ellison, Torrey, Brown, Hani, McAlpine, Moreno, \& Woo}]{patton2020interacting}
Patton, D.~R., Wilson, K.~D., Metrow, C.~J., {et~al.} 2020, Monthly Notices of the Royal Astronomical Society, 494, 4969

\bibitem[{Pettini \& Pagel(2004)}]{pettini2004iii}
Pettini, M., \& Pagel, B.~E. 2004, Monthly Notices of the Royal Astronomical Society, 348, L59

\bibitem[{Ranalli {et~al.}(2003)Ranalli, Comastri, \& Setti}]{ranalli20032}
Ranalli, P., Comastri, A., \& Setti, G. 2003, Astronomy \& Astrophysics, 399, 39

\bibitem[{Rieke {et~al.}(2024)Rieke, Alberts, Shivaei, Lyu, Willmer, P{\'e}rez-Gonz{\'a}lez, \& Williams}]{rieke2024smiles}
Rieke, G., Alberts, S., Shivaei, I., {et~al.} 2024, The Astrophysical Journal, 975, 83

\bibitem[{Rieke {et~al.}(2009)Rieke, Alonso-Herrero, Weiner, P{\'e}rez-Gonz{\'a}lez, Blaylock, Donley, \& Marcillac}]{rieke2009determining}
Rieke, G.~H., Alonso-Herrero, A., Weiner, B., {et~al.} 2009, The Astrophysical Journal, 692, 556

\bibitem[{Rieke {et~al.}(2023)Rieke, Robertson, Tacchella, Hainline, Johnson, Hausen, Ji, Willmer, Eisenstein, Pusk{\'a}s, {et~al.}}]{rieke2023jades}
Rieke, M.~J., Robertson, B., Tacchella, S., {et~al.} 2023, The Astrophysical Journal Supplement Series, 269, 16

\bibitem[{Rujopakarn {et~al.}(2016)Rujopakarn, Dunlop, Rieke, Ivison, Cibinel, Nyland, Jagannathan, Silverman, Alexander, Biggs, {et~al.}}]{rujopakarn2016vla}
Rujopakarn, W., Dunlop, J., Rieke, G.~H., {et~al.} 2016, The Astrophysical Journal, 833, 12

\bibitem[{Shapley {et~al.}(2023)Shapley, Sanders, Reddy, Topping, \& Brammer}]{shapley2023jwst}
Shapley, A.~E., Sanders, R.~L., Reddy, N.~A., Topping, M.~W., \& Brammer, G.~B. 2023, The Astrophysical Journal, 954, 157

\bibitem[{Shtykovskiy \& Gilfanov(2007)}]{shtykovskiy2007high}
Shtykovskiy, P., \& Gilfanov, M. 2007, Astronomy Letters, 33, 437

\bibitem[{Speagle(2020)}]{speagle2020dynesty}
Speagle, J.~S. 2020, Monthly Notices of the Royal Astronomical Society, 493, 3132

\bibitem[{Tetarenko {et~al.}(2020)Tetarenko, Rosolowsky, Miller-Jones, \& Sivakoff}]{tetarenko2020jet}
Tetarenko, A., Rosolowsky, E., Miller-Jones, J., \& Sivakoff, G. 2020, Monthly Notices of the Royal Astronomical Society, 497, 3504

\bibitem[{Tetarenko {et~al.}(2018)Tetarenko, Freeman, Rosolowsky, Miller-Jones, \& Sivakoff}]{tetarenko2018mapping}
Tetarenko, A.~J., Freeman, P., Rosolowsky, E., Miller-Jones, J.~C., \& Sivakoff, G.~R. 2018, Monthly Notices of the Royal Astronomical Society, 475, 448

\bibitem[{Tremmel {et~al.}(2013)Tremmel, Fragos, Lehmer, Tzanavaris, Belczynski, Kalogera, Basu-Zych, Farr, Hornschemeier, Jenkins, {et~al.}}]{tremmel2013modeling}
Tremmel, M., Fragos, T., Lehmer, B., {et~al.} 2013, The Astrophysical Journal, 766, 19

\bibitem[{Veilleux \& Osterbrock(1987)}]{veilleux1987spectral}
Veilleux, S., \& Osterbrock, D.~E. 1987, Astrophysical Journal Supplement Series (ISSN 0067-0049), vol. 63, Feb. 1987, p. 295-310. NSERC-supported research., 63, 295

\bibitem[{Virtanen {et~al.}(2020)Virtanen, Gommers, Oliphant, Haberland, Reddy, Cournapeau, Burovski, Peterson, Weckesser, Bright, {et~al.}}]{virtanen2020scipy}
Virtanen, P., Gommers, R., Oliphant, T.~E., {et~al.} 2020, Nature methods, 17, 261

\bibitem[{Wang {et~al.}(2024)Wang, Leja, de~Graaff, Brammer, Weibel, van Dokkum, Baggen, Suess, Greene, Bezanson, {et~al.}}]{wang2024rubies}
Wang, B., Leja, J., de~Graaff, A., {et~al.} 2024, arXiv preprint arXiv:2405.01473

\bibitem[{Zhang {et~al.}(2012)Zhang, Gilfanov, \& Bogd{\'a}n}]{zhang2012dependence}
Zhang, Z., Gilfanov, M., \& Bogd{\'a}n, {\'A}. 2012, Astronomy \& Astrophysics, 546, A36

\end{thebibliography}
\bibliographystyle{aasjournal}



\end{document}